\documentclass[a4paper,fleqn,usenatbib]{mnras}

\usepackage{newtxtext, newtxmath}
\usepackage{verbatim}
\usepackage[T1]{fontenc}
\usepackage{ae,aecompl}
\usepackage{epsf}

\usepackage{graphicx}	
\usepackage{amsmath}	
\usepackage{amssymb}	
\usepackage{mathtools}
\usepackage{url}
\usepackage{float}
\usepackage{ragged2e}
\usepackage{subcaption}

\newcommand{\dispdot}[2][.4ex]{\dot{\raisebox{0pt}[\dimexpr\height+#1][\depth]{$#2$}}}

\title[X-ray variability of 47TucW]{On the vanishing orbital X-ray variability of the eclipsing binary millisecond pulsar 47 Tuc W}

\author[Hebbar et al.]{
P. R. Hebbar,$^{1}$\thanks{E-mail: hebbar@ualberta.ca}
C. O. Heinke,$^{1}$
D. Kandel,$^{2}$
R. W. Romani,$^{2}$
\newauthor P. C. C. Freire $^{3}$
\\
$^{1}$Department of Physics, CCIS 4-183, University of Alberta, Edmonton, AB, T6G 2E1, Canada\\
$^{2}$Department of Physics, Stanford University, Stanford, CA, 94305, USA\\
$^{3}$Max-Planck-Institut f\"{u}r Radioastronomie, Auf dem H\"{u}gel 69, D-53121 Bonn, Germany
}

\date{Accepted XXX. Received YYY; in original form ZZZ}

\pubyear{2020}

\begin{document}
\label{firstpage}
\pagerange{\pageref{firstpage}--\pageref{lastpage}}
\maketitle

\begin{abstract}
Redback millisecond pulsars (MSPs) typically show pronounced orbital variability in their X-ray emission due to our changing view of the intrabinary shock (IBS) between the pulsar wind and stellar wind from the companion. Some redbacks (``transitional'' MSPs) have shown dramatic changes in their multiwavelength properties, indicating a transition from a radio pulsar state to an accretion-powered state. The redback MSP 47 Tuc W showed clear X-ray orbital variability in the {\it Chandra} ACIS-S observations in 2002, which were not detectable in the longer {\it Chandra} HRC-S observations in 2005--06, suggesting that it might have undergone a state transition. However, the {\it Chandra} observations of 47 Tuc in 2014--15 show similar X-ray orbital variability as in 2002. We explain the different X-ray light-curves from these epochs in terms of two components of the X-ray spectrum (soft X-rays from the pulsar, vs. harder X-rays from the IBS), and different sensitivities of the X-ray instruments observing in each epoch. However, when we use our best-fit spectra with HRC response files to model the HRC light-curve, we expect a more significant and shorter dip than that observed in the 2005--06 {\it Chandra} data. This suggests an intrinsic change in the IBS of the system. We use the ICARUS stellar modelling software, including calculations of heating by an IBS, to model the X-ray,  optical, and UV light-curves of 47 Tuc W. Our best-fitting parameters point towards a high-inclination system ($i \sim 60 \deg$), which is primarily heated by the pulsar radiation, with an IBS dominated by the companion wind momentum. 
\end{abstract}




\section{Introduction}

Shortly after the discovery of the first MSP, PSR B1937+21 \citep{backer1982}, it was proposed that MSPs are recycled NSs, spun up by accretion from a companion star during an X-ray binary phase \citep{alpar1982,bhattacharya1991}. Support for this evolutionary model has come from detections of millisecond X-ray pulsations in many accreting NSs \citep{wijnands1998,Patruno12}; and the observed transitions of IGR J18245-2452 \citep{papitto2013, Campana_2018}, PSR J1023+0038 \citep[etc.]{archibald2009,stappers2014,tendulkar2014} and XSS J12270-4859 \citep{deMartino10,Hill11,Bassa14,Bogdanov14,Roy15} between an X-ray bright ($L_X > 10^{36}$ erg/s) low-mass X-ray binary (LMXB) state (observed only in IGR J18245-2542); a state of intermediate X-ray brightness ($L_X \sim10^{33-34}$ erg/s) with X-ray pulsations and evidence of a  disk, suggesting low-level accretion; and a radio pulsar state. Detailed studies of PSR J1023+0038 across multiple wavelengths reveal that the X-ray intermediate state  consists of three ``modes" \citep{Archibald15,bogdanov2015,Ambrosino17,Papitto2019} - a high flux mode ($\sim 10^{33}$ ergs s$^{-1}$) during which coherent X-ray and optical pulsations are observed (perhaps indicating active accretion); a low flux mode ($\sim 10^{32}$ ergs s$^{-1}$) where no X-ray pulsations have been observed, probably due to the accretion flow being pushed away from the pulsar by the pulsar wind; and sporadic X-ray and UV flares reaching up to $\sim 10^{34}$ ergs s$^{-1}$. Similar X-ray mode-switching behaviour is seen in the other two verified transitional MSPs \citep{papitto2013,deMartino13} and in another two candidate systems \citep{BogdanovHalpern15,CotiZelati19}. In contrast, there is no accretion during the radio pulsar state. The X-ray emission during this state consists of a dominant shock component, that shows double-peaked orbital modulation, and sometimes a fainter, softer component, likely from heated magnetic polar caps \citep[e.g.][]{bogdanov2005,Bogdanov10,bogdanov11b,hui15}.

As predicted by the recycling model for formation of MSPs, the majority of them are found in binary systems. While only $\sim 2\%$ of longer-period radio pulsars are found in binary systems \citep{lyne1990}, $\sim 80\%$ of MSPs have a companion star \citep{camilo2005}.  A significant fraction of these binary MSPs show radio eclipses, in which radio pulsations cannot be detected during a fraction of the orbit, typically around the superior conjunction of the NS. These eclipsing binary MSPs are classified into two groups - black widows with tiny, partly degenerate companions of mass $M_2 << 0.1M_{\odot}$, and the larger redbacks, with $M_2 \sim 0.1 - 0.4M_{\odot}$ \citep{freire2005,roberts2011}. Redbacks have companions that are non-degenerate low-mass main sequence or subgiant stars. The first few redback MSPs were discovered in globular clusters (GCs; \citealt{camilo2000,d'Amico01}) \footnote{Refer http://www.naic.edu/$\sim$pfreire/GCpsr.html for up-to-date catalog of MSPs}, but targeted observations of the error circles of Fermi gamma-ray sources have identified many of these in the galactic field \citep[e.g.][]{Roberts13}.

\subsection{Intra-binary shock and irradiation feedback}
Irradiation feedback from the pulsar heats up the companion and causes companion mass loss  \citep[quasi-Roche lobe overflow;][]{benvenuto14, benvenuto15}. This stellar wind driven through pulsar heating can interact with the relativistic pulsar wind forming an intra-binary shock \citep[IBS,][]{bednarek14}. The structure of the IBS depends on momentum balance between the pulsar and stellar winds \citep[e.g.][]{wadiasingh2018}. Electrons accelerated in this shock cool through synchrotron or inverse Compton processes. The X-ray light-curve of a typical eclipsing binary pulsar shows a double peaked structure due to Doppler beaming of the IBS radiation. Such a light-curve is sensitive to the binary parameters of the system, thus its careful modelling allows us to constrain the system parameters. Early models of the shock considered only the pulsar wind dominated scenario where the prominent dip in X-ray flux near the radio eclipse was explained as the occultation of the shock close to the L1 point \citep[e.g.][]{bogdanov2005}. However, detailed modelling of the IBS has shown that Doppler beaming of an X-ray emitting shock in the companion wind dominated case can also cause a significant drop in the X-ray flux during the radio eclipse \citep{romani2016, sanchez2017, li2014, wadiasingh2018, alnoori2018}.  

The companions in compact tidally locked redback and black widow systems have strong day-night heating asymmetry. The resulting temperature difference across the surface of the companion leads to variability in the optical bands \citep[etc.]{Fruchter88b,bogdanov11b,Bogdanov14, romani2011, kong2012, breton2013}. Given the orbital ephemeris from radio and $\gamma$-ray observations, studying the optical and X-ray light-curves could reveal the details of pulsar heating mechanisms, companion wind, component masses and the geometry of the systems \citep{djorgovisky88, callanan1995}. The detailed modelling of these light-curves delivers an inclination of the system, a key ingredient for  precise mass measurements of pulsars. Such mass measurements are crucial  to constraining the NS equation of state \citep[e.g.][]{vankerkwijk2011}.  

\subsection{47 Tuc W}
47 Tuc hosts 25 known pulsars, all MSPs with $P\sim$ 2--8 ms,  of which 15 are in binary systems \citep{Manchester91,camilo2000,ridolfi2016,Pan2016,freire2017}. Radio eclipses have been detected in five of these pulsars, of which two are confirmed to be redbacks - PSR J0024-7204W (47 Tuc W), and PSR J0024-7201V (47 Tuc V). 47 Tuc W was first detected by \citet{camilo2000} using the Parkes radio telescope, with a period of 2.35 ms and an orbital period of 3.2 hrs. The pulsar was eclipsed in the  radio for $\sim$ 25\% of its orbit. The position and the nature of the companion were deduced by identifying a periodic variable in a long series of Hubble Space Telescope (HST) exposures, which showed variability matching the pulsar's known orbital period \citep{edmonds2002}.  47 Tuc W has a  companion consistent with a main sequence star, with $M_2 > 0.13 M_{\odot}$ \citep{edmonds2002}. The most up-to-date timing solution for this system gives an orbital frequency $f_b = 8.71 \times 10^{-5}$ s$^{-1}$, but multiple orbital period derivatives are necessary to find a timing solution \citep{ridolfi2016}. 

The position of 47 Tuc W is coincident with that of X-ray source W29 detected by Chandra ACIS observations of the core of 47 Tuc \citep{grindlay2001,heinke2005}. X-ray analysis of 47 Tuc W showed that the emission consisted of two components - a hard non-thermal component which contributes about 70\% of the observed X-ray luminosity, and shows a decrease in flux for $\sim$ 30\% of the orbit; and a soft, thermal component, which did not show variability \citep{bogdanov2005}. \citet{bogdanov2005} proposed that the non-thermal spectrum is from an IBS due to interaction of the pulsar wind with the stellar wind, and that the thermal component arises from the neutron star surface, as seen in other MSPs  \citep[e.g.][]{Becker93,Zavlin02}. Observations of 47 Tuc W using the Chandra HRC-S instrument in 2005--06 indicated an absence of X-ray eclipses, although the total exposure, and number of collected counts, were larger \citep{cameron2007}. This suggested the possibility that  47 Tuc W, like several other redbacks, may be a transitional MSP, and may have engaged in a state transition between these observations.

In this paper, we use the 2014--15 Chandra ACIS-S observations of 47 Tuc  \citep[PI  Bogdanov;][]{bogdanov2016,Bahramian2017,bhattacharya2017}, along with the previous ACIS-S \citep[PI Grindlay;][]{heinke2005} and HRC-S \citep[PI  Rutledge;][]{cameron2007} data to look into this puzzle. We also perform phase-resolved X-ray spectroscopy on the ACIS-S data to look into any changes in the spectra between the two ACIS-S   epochs, and to derive the properties of the suggested IBS. In Section~\ref{sec: obs}, we give a detailed description of how we extracted the source photons and analysed the data. Section~\ref{sec: variability} discusses our analysis of the X-ray light-curve of 47 Tuc W, and explains our hypothesis for why variability was not observed in the HRC data. We analyse the X-ray spectra of the system in Section~\ref{sec: spectra}, and use this model to verify the HRC light-curves. In Section~\ref{sec:optical}, we use the ICARUS modules developed by \citet{Kandel19} to model the optical \citep{edmonds2002,bogdanov2005} and X-ray light-curves of 47 Tuc W and study the properties of the IBS and its heating of the companion. We summarize and conclude our findings in Section~\ref{sec:conclusions}.

\begin{figure*}
    \centering
    \includegraphics[width=0.8\textwidth]{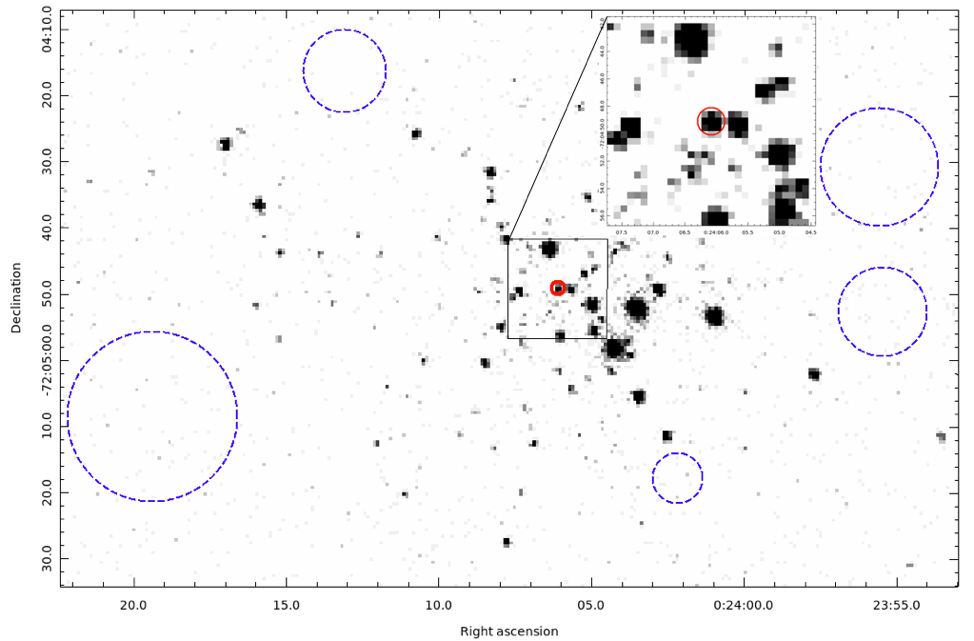}
    \caption{X-ray image of the globular cluster 47 Tuc. The solid red circle of radius $1 \arcsec$ centred at $\alpha = 00^{\mathrm{h}} 24^{\mathrm{m}} 06 \fs 1, \delta = -72^{\circ} 04 \arcmin 49 \farcs 1$ shows the region from which source photons of 47 Tuc W were extracted. We select 5 regions around 47 Tuc  without any X-ray sources for background extraction (shown in dashed blue).}
    \label{fig:regions}
\end{figure*}

\section{Observations and Data reduction}
\label{sec: obs}
Due to crowding in the compact core of 47 Tuc, we used the sub-arcsecond resolution of the  {\it Chandra X-ray Observatory} and the  {\it Hubble Space Telescope} to investigate 47 Tuc W. 

\subsection{X-ray observations}

The Chandra X-ray Observatory uses two types of X-ray detectors  \footnote{http://cxc.harvard.edu/proposer/POG/}. Chandra's ACIS detectors are charge-coupled devices (CCDs), where absorption of an X-ray photon liberates a proportional number of electrons from a pixel of a  semiconductor chip. The ACIS detectors are generally operated to observe for an exposure time (typically 3.2 seconds), then transfer the charges to readout electronics. The ACIS detectors initially had high sensitivity to X-rays both above and below 1 keV, but have lost much of their lower-energy sensitivity due to an increasing layer of absorbing contaminant \footnote{http://cxc.harvard.edu/ciao/why/acisqecontamN0010.html}. Chandra's HRC detectors, on the other hand, are comprised of microchannel plates, where X-rays liberate electrons, which are accelerated by an applied voltage down the microchannels to produce an avalanche of charge at the readout. The HRC detectors have lower sensitivity than ACIS overall, but are now more sensitive than ACIS below 1 keV. Thus, the ACIS detectors retain (limited) spectral information, have higher sensitivity above 1 keV, and have very low background, while the HRC detectors retain sub-millisecond timing information, have slightly higher angular resolution, and have higher sensitivity below 1 keV. 

We analysed the ACIS-S 2002 observations (exposure $\sim 300$ks) and 2014--15 observations (exposure $\sim 200$ks) to construct the X-ray spectra and the light-curves of the 47 Tuc W system. We also studied the HRC observations of 2005--06 (exposure $\sim 800$ ks) to verify the light-curves presented by \citet{cameron2007} and to check for rotational variability in 47 Tuc W using the new orbital ephemeris of \citet{ridolfi2016}.
The details of all X-ray observations used are summarised in Table \ref{table:data}. 

\begin{table*}
\centering
\caption{Summary of X-ray observations used}
\label{table:data}
\begin{tabular}{cccc|ccccc}
\hline
Obs ID & Instrument & Exposure(ks) & Start Date          &  & Obs ID & Instrument & Exposure(ks) & Start Date          \\ \hline
2735   & ACIS-S     & 65.24        & 2002-09-29 16:57:56 &  & 5542   & HRC-S      & 49.76        & 2005-12-19 07:03:06 \\
3384   & ACIS-S     & 5.31         & 2002-09-30 11:37:18 &  & 5543   & HRC-S      & 50.65        & 2005-12-20 14:57:42 \\
2736   & ACIS-S     & 65.24        & 2002-09-30 13:24:28 &  & 5544   & HRC-S      & 49.83        & 2005-12-21 23:25:20 \\
3385   & ACIS-S     & 5.31         & 2002-10-01 08:12:28 &  & 5545   & HRC-S      & 51.64        & 2005-12-23 05:01:54 \\
2737   & ACIS-S     & 65.24        & 2002-10-02 18:50:07 &  & 5546   & HRC-S      & 48.27        & 2005-12-27 05:33:43 \\
3386   & ACIS-S     & 5.54         & 2002-10-03 13:37:18 &  & 6230   & HRC-S      & 44.77        & 2005-12-28 13:44:36 \\
2738   & ACIS-S     & 68.77        & 2002-10-11 01:41:55 &  & 6231   & HRC-S      & 46.89        & 2005-12-29 21:50:23 \\
3387   & ACIS-S     & 5.73         & 2002-10-11 21:22:09 &  & 6232   & HRC-S      & 44.15        & 2005-12-31 05:17:20 \\
15747  & ACIS-S     & 50.04        & 2014-09-09 19:32:57 &  & 6233   & HRC-S      & 97.18        & 2006-01-02 05:37:31 \\
15748  & ACIS-S     & 16.24        & 2014-10-02 06:17:00 &  & 6235   & HRC-S      & 49.93        & 2006-01-04 04:04:57 \\
16527  & ACIS-S     & 40.88        & 2014-09-05 04:38:37 &  & 6236   & HRC-S      & 51.7         & 2006-01-05 11:29:07 \\
16528  & ACIS-S     & 40.28        & 2015-02-02 14:23:34 &  & 6237   & HRC-S      & 49.96        & 2005-12-24 14:07:36 \\
16529  & ACIS-S     & 24.7         & 2014-09-21 07:55:51 &  & 6238   & HRC-S      & 48.2         & 2005-12-25 21:12:00 \\
17420  & ACIS-S     & 9.13         & 2014-09-30 22:56:03 &  & 6239   & HRC-S      & 49.88        & 2006-01-06 22:08:49 \\
       &            &              &                     &  & 6240   & HRC-S      & 49.07        & 2006-01-08 02:19:31 \\ \hline
\end{tabular}

\end{table*}

CIAO version $4.9$ was used for data reduction and image processing. The initial data downloaded from WebChaSeR was reprocessed  according to CALDB $4.7.6$ calibration standards using the \texttt{chandra\_repro} command. The parameters \texttt{`badpixel'} and \texttt{`process\_events'} were set to ``yes" in order to create new level=1 event and badpixel files using the latest calibrations. In order to prevent good events from being removed, the parameter \texttt{`check\_vf\_pha'} was set to ``no". The \texttt{`pixel\_adj'} parameter was set to ``default" (EDSER) in order to obtain the maximal spatial resolution. The X-ray photons corresponding to  $47$ Tuc W were extracted from a circular region of $1"$ radius around the source ($\alpha = 00^{\mathrm{h}} 24^{\mathrm{m}} 06 \fs 1; \delta = -72^{\circ} 04 \arcmin 49 \farcs 1$) as shown in Fig.~\ref{fig:regions}. This region was selected as a compromise between selecting maximum photons from the target, and avoiding photons from the neighbouring source W32. Photons were corrected for barycentric shifts using the \texttt{axbary} tool of CIAO. The aspect solution file and the exposure statistics file were also corrected in order to correct the good time intervals. Five regions close to  $47$ Tuc, and free of sources, were chosen for background subtraction, and barycentric correction was also applied to the background light-curves.

\subsection{Optical Observations}
We use the reduced, folded optical and UV data as presented in \citet{bogdanov2005} and \citet{cadelano2015} for our analysis. {\it Hubble Space Telescope} observations of 47 Tuc reveal the optical light-curve of the companion star. \citet{bogdanov2005} plot the optical light-curves in the ACS bands F435W (67 points), F475W (19 points), F555W (8 points) and F606W (6 points), but only as normalised fluxes as a function of an unpublished radio ephemeris. However, they also plot fluxes in these four bands with the SED of the companion at maximum, and from this we can estimate the fluxes of the individual detections.  \citet{cadelano2015} plotted WFC3 F300X (12 points) and F390W (5 points) magnitudes (on the HST system) as a function of binary phase on an updated radio ephemeris by P. Freire. We direct readers to the above mentioned papers for details of the optical data reduction methods.

\subsection{X-ray orbital and spin variability analyses}
\begin{figure}
\centering
\includegraphics[width=0.47\textwidth]{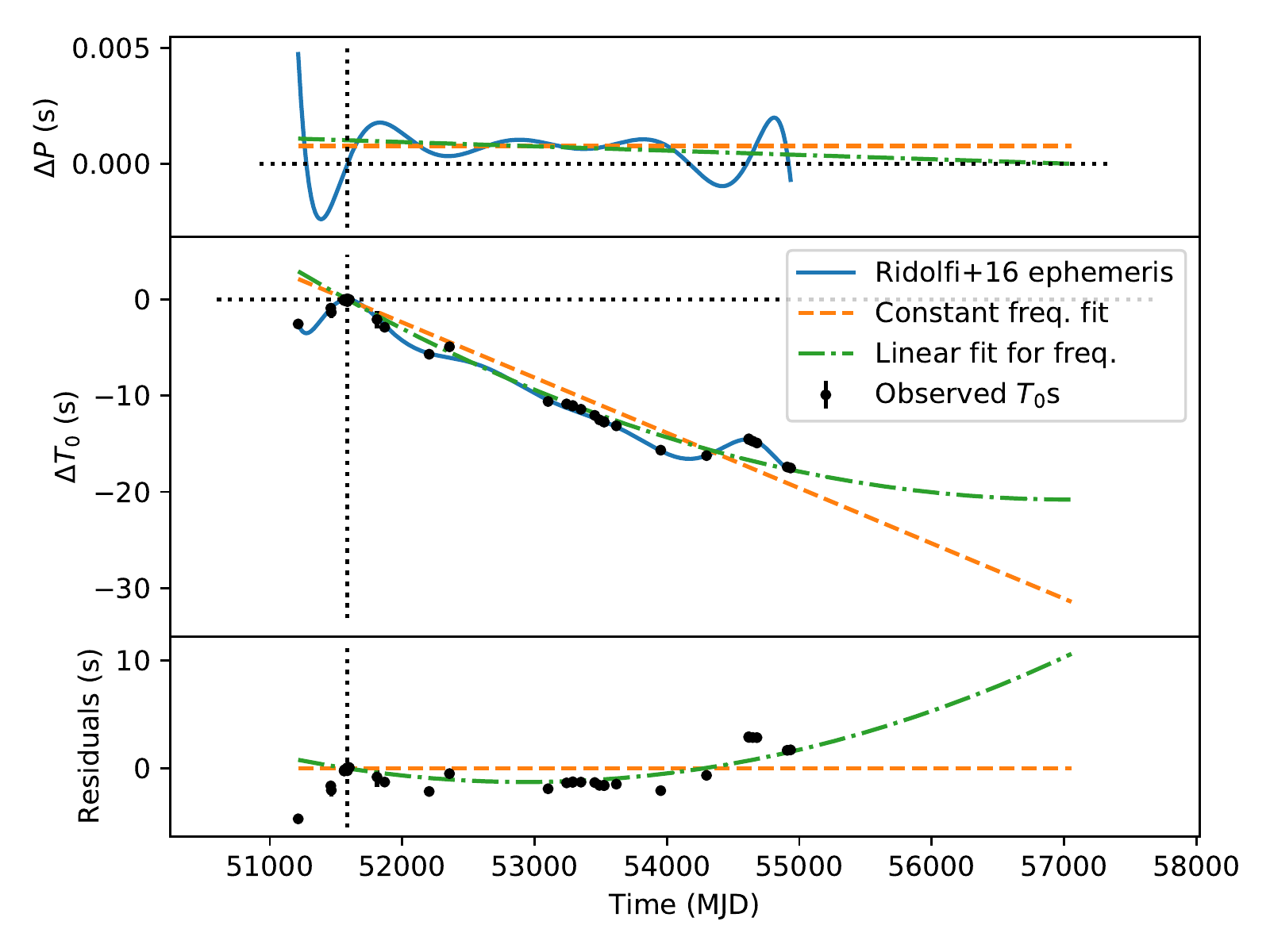}
\caption{Variation in orbital parameters of 47Tuc W. The vertical dotted line represents the time of periapsis passage mentioned in \citet{ridolfi2016} and used as a reference epoch for their orbital ephemeris solution. ({\it Top}:) Change in the orbital period of 47 Tuc W. ({\it Middle}:) Deviation of observed times of the periapsis passage from the times predicted assuming a Keplerian orbit. In both the plots the solid blue line represents the \citet{ridolfi2016} solution. In order to predict the change in $\Delta T_0$ for the duration of the 2014--15 observation, we fit the observed $\Delta T_0$s under a constant frequency (shown dashed orange), and a linearly changing frequency with time (i.e. constant $\dispdot{f}$; shown in dotted-dash green) assumptions. We note that the extrapolated $\Delta T_0$ is $\sim$ 20--30s, which is much smaller than the width of our bins ($\sim 1000$s). ({\it Bottom}:) Residuals with respect to the constant frequency fit. Though the size of residuals are much larger than the errorbars, they are smaller than our binning time.}
\label{fig:ephem}
\end{figure}

We used the ephemeris data of 47 Tuc W  from \citet{ridolfi2016} to prepare the phase folded X-ray light-curve. However, this ephemeris contains many derivatives, and thus diverges for times (such as our 2015 epoch) outside the epochs over which it is defined (February 5, 1999 to April 13, 2009). However, in Fig.~\ref{fig:ephem} (plotted from February 5, 1999 to February 3, 2015), we see that the change in time of passage of periapsis changes by $\lesssim 20$ s across $\sim 11$ years (the duration over which the radio pulses were observed). Extrapolating this to the epoch of our 2015 data, we see that the change in $\Delta T_0 \sim 20-30$ s, which is much less than the time interval corresponding to our phase bins ($\sim 1000$ s). Thus we use the orbital frequency of \citet{ridolfi2016} without any modification.

We then constructed phase-folded light-curves in the energy interval $0.3-8.0$\,keV separately for each observation from the corresponding barycentre-corrected source and background files using the  and \texttt{dmextract} tool. To verify our methods, we compare our 2002 light-curve with that of Fig. 1 in \citet{bogdanov2005}. We find that the primary difference between the two light-curves is the position of the phase 0 epoch; \citet{bogdanov2005} uses inferior conjunction of the pulsar as the phase 0 epoch, while \citet{ridolfi2016} uses the time of perihelion passage of the pulsar as the phase 0 epoch. After shifting the light-curves appropriately along the x-axis, the two light-curves are consistent within their error bars ($\chi^2/$d.o.f = 2.0/19). These residual differences between the two light-curves could be due to the updated versions of \textsc{CIAO}, and \textsc{CALDB}, used in this paper. 

We created phase-folded light-curves for observations taken in 2002, 2005--06 and 2014--15 separately. This enables us to study any possible differences in the light-curves across the three epochs. We binned the light-curves to intervals of 0.1 in phase so that the counts per bin are above 10 for most bins, so that Pearson's $\chi^2$ statistical test may be used \citep{Pearson_1900}. We shall use Pearson's definition of $\chi^2$ for our X-ray light-curve analysis. Since a few bins have net counts smaller than 10, we use Gehrels' error bars \citet{Gehrels_1986} for plotting purposes alone. In order to study the variability across different energy bands, we constructed additional light-curves in a very low energy band (0.2 - 1.0\,keV) and a high energy band (2.0 - 8.0\,keV) for the ACIS-S 2002 and 2014--15 observations. For the light-curves in very low energy and high energy bands, we needed a larger bin size of 0.2 to have reasonable counts in each bin.  In order to study the IBS in more detail, we also constructed light-curves with bin size 0.06 to show the dip due to Doppler beaming more clearly.

We also use the \citet{ridolfi2016} updated ephemeris solution to fold the HRC data and search for millisecond, spin-period pulsations. We calculate the phase of each photon after accounting for R\"{o}mer delays due to orbital motion of the MSP. We then check for variability using the $Z_n^2$-test \citep{buccheri1983}, where the optimum $n$ was chosen using the $H$-test \citep{dejager1989}. We observe a $Z_n^2$ value 3.05 for $n = 1$. This shows variability of $< 2\sigma$ significance; i.e., the rotational phase-folded HRC light-curves do not show clear evidence for variability.

\subsection{Spectral analysis}

The HRC observations do not contain sufficient information regarding the energy of the photons and hence cannot be used for spectral analysis. We analysed the data from the 2002 and 2014--15 ACIS observations separately to investigate any change in the spectrum. To look into the nature of the dips in the light-curves, we grouped the data from the phase intervals 0.1--0.4, when the $47$ Tuc W system showed a decrease in the count rate, separately from the remaining intervals. For this purpose, we extracted the good time intervals corresponding to these phases using the \texttt{dmgti} tool, and aligned them to the corresponding ACIS observations using the \texttt{gti\_align} tool. We then extracted the spectrum from individual observations 
using the \texttt{spec\_extract} tool and  combined them using \texttt{combine\_spectra} in order to increase
the photon count. We grouped the spectra such that each bin had at least 1 photon, and used C-statistics \citep{Cash1979} for spectral analysis. $\chi^2$ statistics with more conservative binning gave similar results, with larger errors. For the purpose of plotting alone, the data points were re-binned such that each bin had at least fifteen counts.

\begin{figure*}
    \centering
    \begin{subfigure}[b]{0.45\textwidth}
        \includegraphics[width=\textwidth]{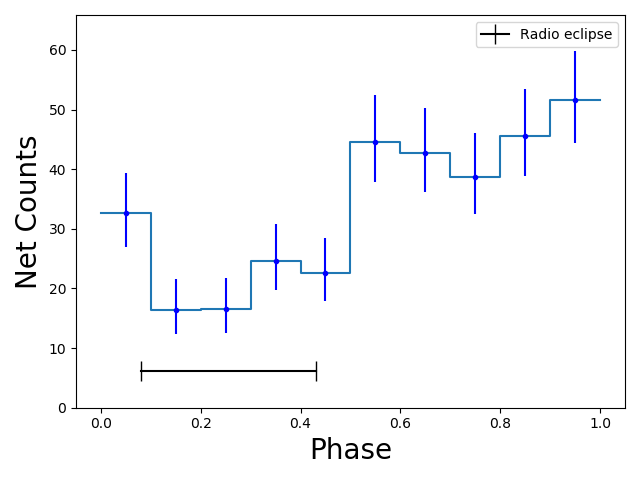}
        \caption{Orbital phase-folded light-curve of 47 Tuc W from ACIS 2002 observations. $\chi^2$/d.o.f = 44.1/9 ($p$-value = $1.4 \times 10^{-6}$).}
        \label{fig:lc_2002}
    \end{subfigure}
    \hfill
    \begin{subfigure}[b]{0.45\textwidth}
        \includegraphics[width=\textwidth]{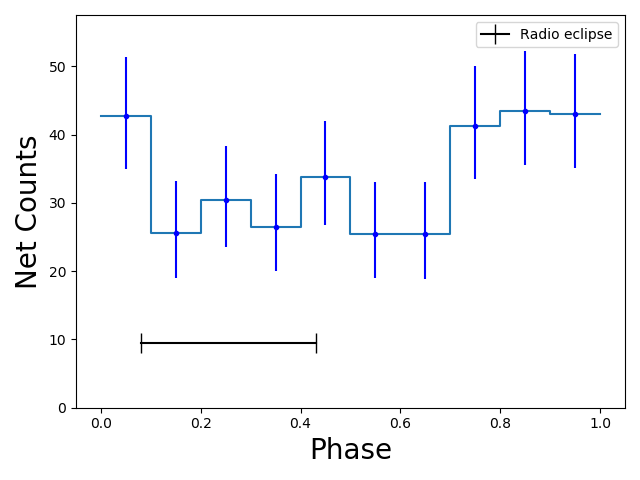}
        \caption{Orbital phase-folded light-curve of 47 Tuc W from HRC 2005--06 observations. $\chi^2$/d.o.f = 17.2/9 ($p$-value = 0.05).}
        \label{fig:lc_2005}
    \end{subfigure}
    
    \begin{subfigure}[b]{0.45\textwidth}
        \includegraphics[width=\textwidth]{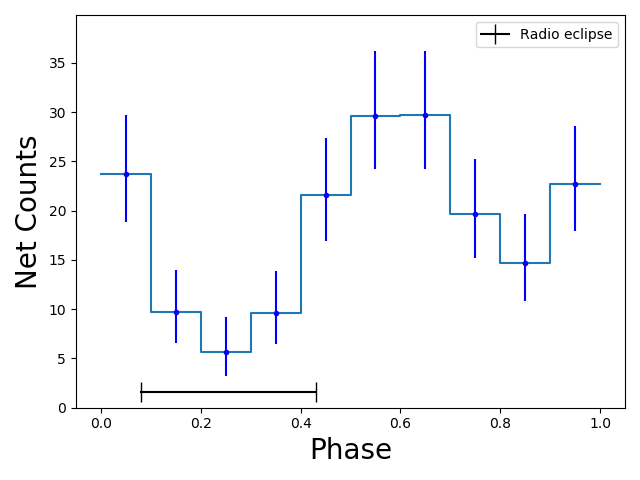}
        \caption{Orbital phase-folded light-curve of 47 Tuc W from ACIS 2014--15 observations. $\chi^2$/d.o.f = 34.4/9 ($p$-value = $7.7 \times 10^{-5}$).}
        \label{fig:lc_2015}
    \end{subfigure}
    \hfill
    \begin{subfigure}[b]{0.45\textwidth}
        \includegraphics[width=\textwidth]{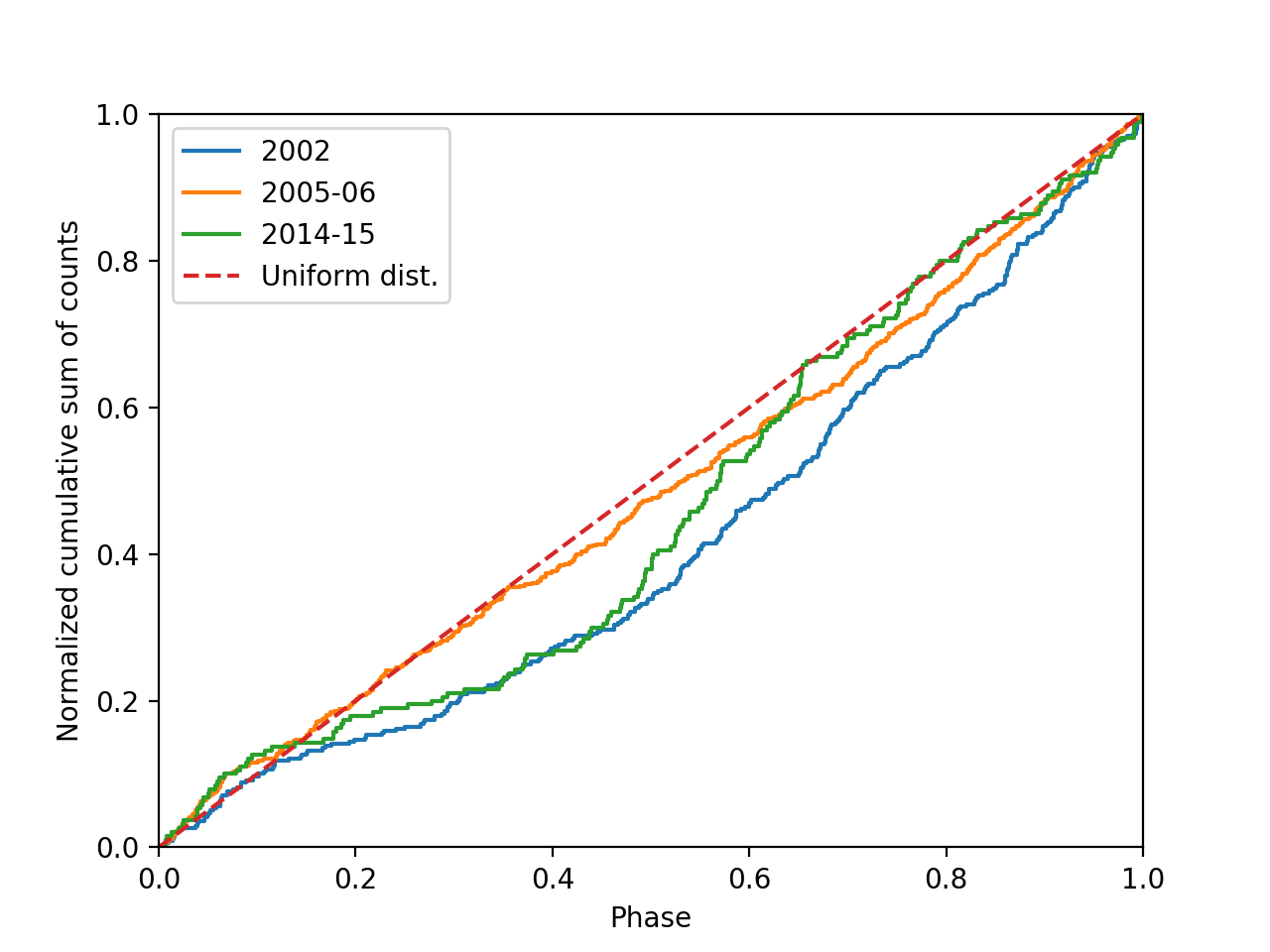}
        \caption{Cumulative distribution of photons from the 2002, 2005--06, and 2014--15 observations. Uniform distribution is also plotted for comparison.}
        \label{fig:lc_cumm}
    \end{subfigure}
    \caption{X-ray light-curve analysis of 47 Tuc W. The solid black line at the bottom of the light-curves shows the position of the radio eclipse (0.09--0.43). We observe that the net count rate in 2002 and 2014--15 ACIS light-curves (extracted in the 0.3--8.0 keV energy range) decreases during phase interval 0.1--0.4, consistent with the radio eclipse. However the light-curve from 2005--06 HRC data does not show variability. We also show the cumulative distribution of the photons during the three observations to give an unbinned perspective and compare them to a uniform distribution (constant net count-rate). Performing the Kolmogorov-Smirnov test  (KS test) between the unbinned light-curves and a uniform distribution gives values 0.17 ($p-$value = $1.2 \times 10^{-8}$) for 2002 data, 0.16 ($p-$value = $1.7 \times 10^{-4}$) for the 2015 data, and 0.063 ($p-$value = 0.032 i.e $< 3\sigma$ confidence) for the 2005 data.}
    \label{fig:X-ray_lc}
\end{figure*}

\section{X-ray Variability}
\label{sec: variability}

\begin{figure*}
\centering
\begin{subfigure}[b]{0.45\textwidth}
 \includegraphics[width=\textwidth]{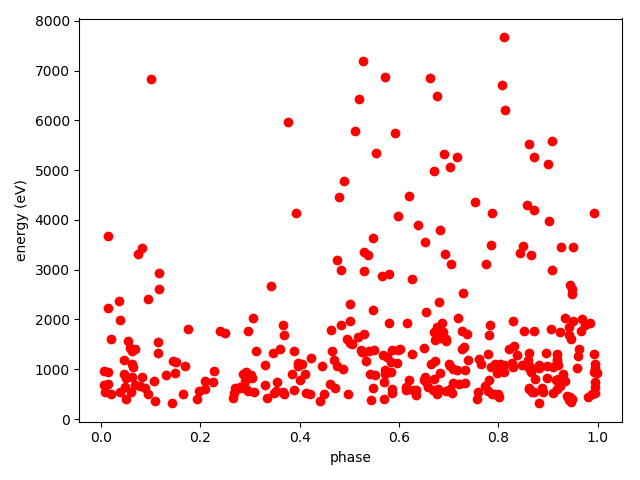}
 \caption{2002 ACIS observations}
 \label{fig:2002enph}
\end{subfigure}
\hfill
\begin{subfigure}[b]{0.45\textwidth}
 \includegraphics[width=\textwidth]{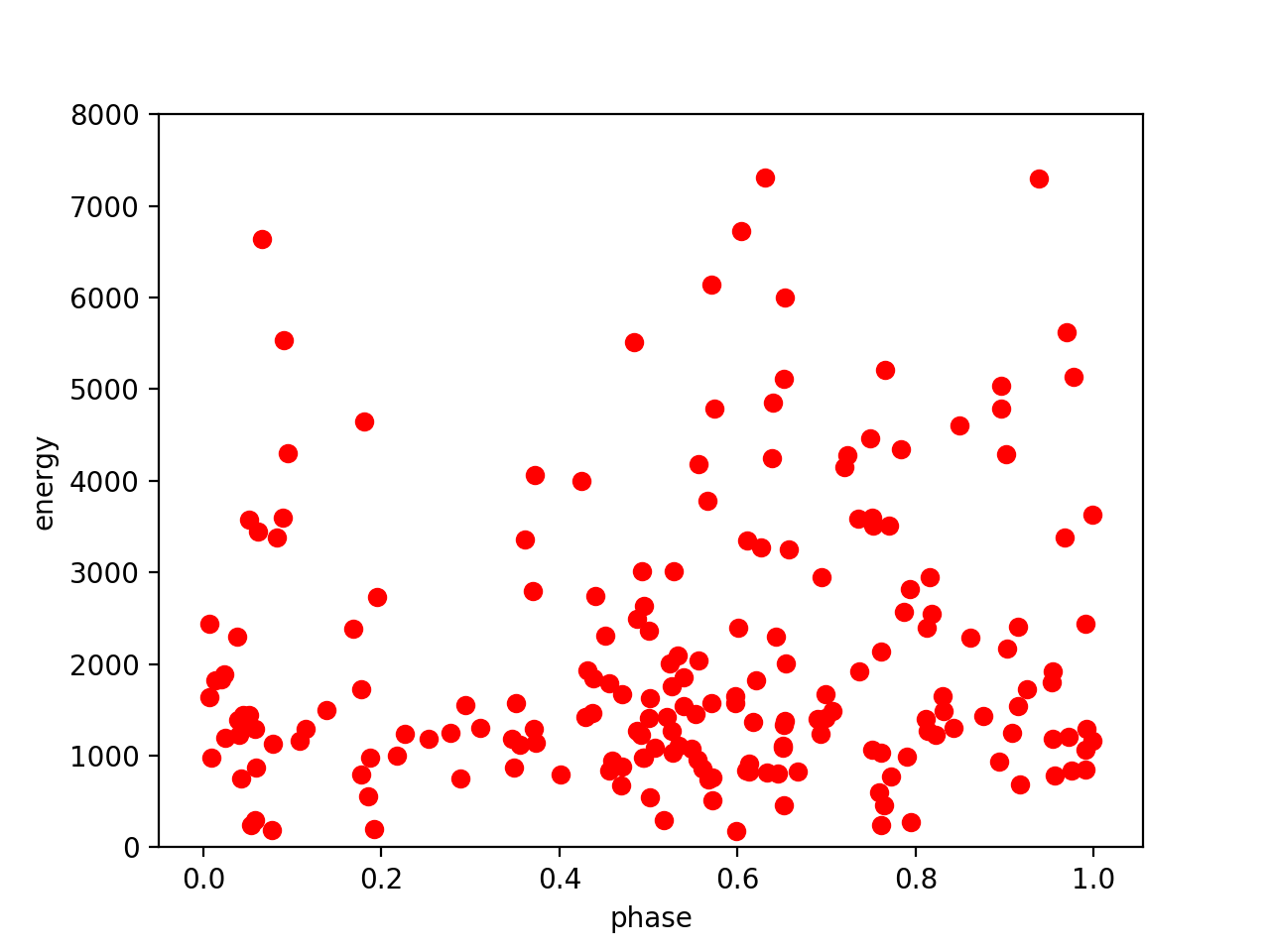}
\caption{2014--15 ACIS observations}
\label{fig:2015enph}
\end{subfigure}
\caption{Scatter plot of photon energy vs. phase for the 2002 and 2014--15 data. Each dot represents an X-ray photon detected. We see that the source exhibits larger variability at higher energies than at lower energies.}
\label{fig:enph}
\end{figure*}

\begin{figure*}
    \centering
    \begin{subfigure}[b]{0.45\textwidth}
     \includegraphics[width=\textwidth]{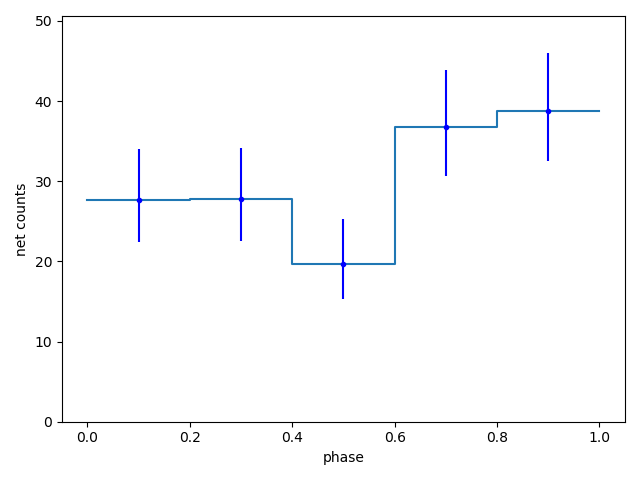}
     \caption{Phase folded light-curve of 2002 ACIS observations in the energy interval 0.2--1.0 keV. $\chi^2$/d.o.f = $7.9/4$ ($p$-value $=0.1$)}
     \label{fig:2002_low}
    \end{subfigure}
    \hfill
    \begin{subfigure}[b]{0.45\textwidth}
     \includegraphics[width=\textwidth]{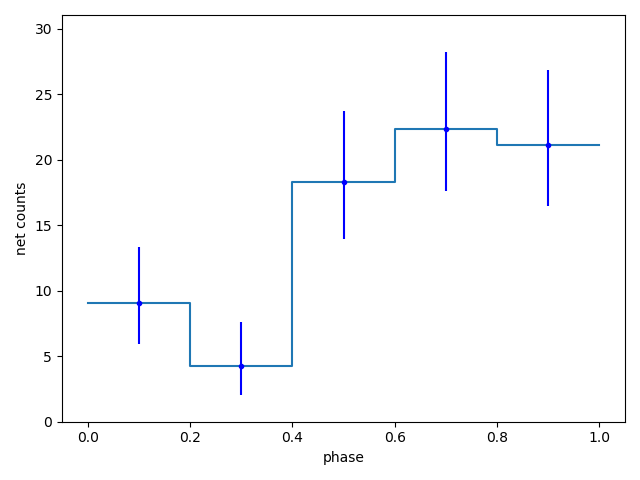}
     \caption{Phase folded light-curve of 2002 ACIS observations in the energy interval 2.0--8.0 keV. $\chi^2$/d.o.f = $16.90/4$ ($p$-value $=0.002$)}
     \label{fig:2002_high}
    \end{subfigure}
    
    \begin{subfigure}[b]{0.45\textwidth}
     \includegraphics[width=\textwidth]{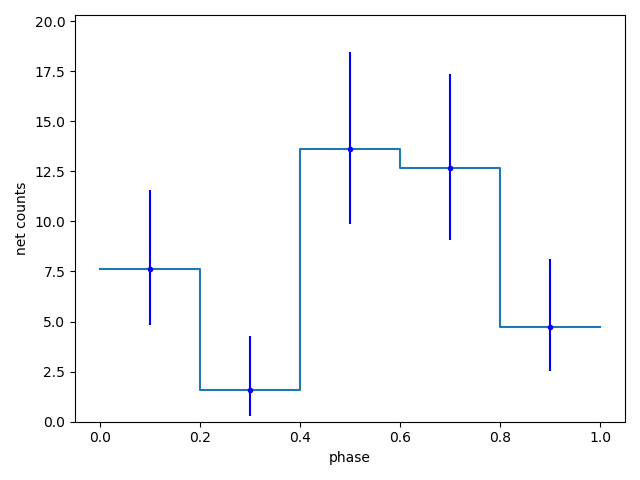}
     \caption{Phase folded light-curve of 2014--15 ACIS observations in the energy interval 0.2--1.0 keV. $\chi^2$/d.o.f = $13.0/4$ ($p$-value $=0.01$)}
     \label{fig:2015_low}
    \end{subfigure}
    \hfill
    \begin{subfigure}[b]{0.45\textwidth}
     \includegraphics[width=\textwidth]{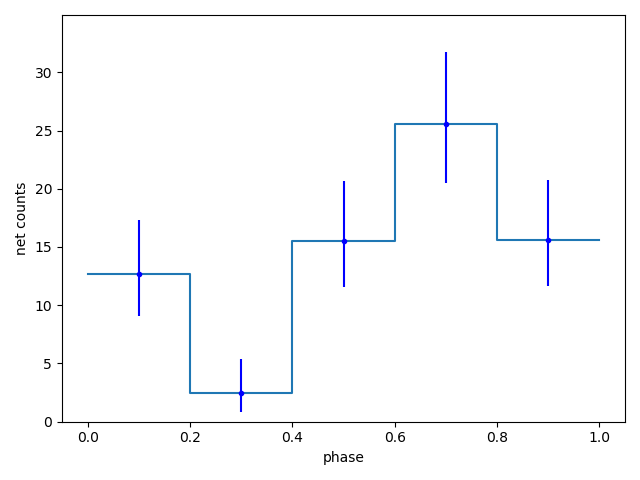}
     \caption{Phase folded light-curve of 2014--15 ACIS observations in the energy interval 2.0--8.0 keV. $\chi^2$/d.o.f = $19.0/4$ ($p$-value $8 \times 10^{-4}$)}
     \label{fig:2015_high}
    \end{subfigure}
    \caption{Phase folded light-curves of 47 Tuc W in low (0.2--1.0 keV) and high (2.0-8.0 keV) and high energies. We see that the variability is more significant at higher energies.}
    \label{fig:my_label}
\end{figure*}

We extracted the light-curves from the 2002 ACIS observations, and the 2005--06 HRC observations, from the reprocessed data, after accounting for the changes in the calibration standards. Figure~\ref{fig:lc_2002} shows the phase-folded light-curve from the 2002 observations. We clearly see the dip in the light-curve identified by \citet{bogdanov2005} during phases $0.1 - 0.5$. These phases correspond to the radio eclipse of the MSP \citep[$0.09 - 0.43$,][]{ridolfi2016}. Fitting the folded light-curve to a constant gives $\chi^2 = 44.1$ for 9 degrees of freedom (d.o.f), giving the probability of the null hypothesis (no variability), $p = 1.4 \times 10^{-6}$. Figure \ref{fig:lc_2005} shows the phase-folded light-curve from the 2005-006 HRC-S observations. As can be seen from the graph, the changes in the net counts per bin are  smaller, and less significant. Fitting this curve to a constant value gives $\chi^2 = 17.2$, i.e $p = 0.05$. Thus this light-curve is consistent (at the 2$\sigma$ level) with a constant source, as noted by \citet{cameron2007}.

Figure \ref{fig:lc_2015} shows the light-curve extracted from the 2014--15 ACIS observations, which shows distinct similarities to that of the 2002 observations, though also some differences. A fit of a constant to the 2014--15 light-curve gives $\chi^2 = 34.4$ for 9 d.o.f. The $\chi^2$ value for 9 d.o.f corresponds to a null hypothesis (that the source has constant count rate) probability of $7.7 \times 10^{-5}$. However, the 2014--15 observation shows a deeper (and narrower) dip at phase $\sim$0.2 than in 2002, and a second, shallower dip at phase $\sim$0.85 not present in 2002. We saw earlier that the error in the phase between 2002 and 2015 observations, $\Delta \phi \leq 0.1$, which corresponds to $\leq$1 phase bin. Thus, errors in the phase are unlikely to explain these differences. Instead the changes are likely due at least in part to the decreased ACIS low energy response in 2015, which decreased the contribution of the unmodulated thermal component.

To visualise the difference in the 2005 light-curve from the other two light-curves, we first plotted a scatter plot of photon energy vs. phase for the 2002 data, as shown in Fig.~\ref{fig:2002enph}. The density of 
points in a region depicts the number of photons observed with the given energy during a given phase. From this figure, it can be seen that the variation in the count rate is much more prominent for high energy photons, in comparison to the lower energy photons which show a much smaller variation. Thus, the lower sensitivity of the HRC detector to the variable, higher energy X-rays might be the reason for the absence of variability. To investigate this, we plotted the light-curve of the 2002 observations for photons in the energy interval 0.2 - 1.0 keV (shown in Fig.~\ref{fig:2002_low}). We increased the bin size to 0.2 so that each bin has a reasonable number of photons to apply Gaussian statistics. This reduces the $\chi^2$/d.o.f to $7.9/4$, i.e. the probability that the source has a constant light-curve in this range is $0.10$ - similar to that of the 2005 light-curve. On the other hand, in figure \ref{fig:2002_high}, the light-curve for high energy photons (2.0 - 8.0 keV) from the ACIS 2002 observations, the change in counts per bin in the light-curve is much more significant, $\chi^2 = 16.90$ for 4 d.o.f, i.e. the probability of the source having a constant count rate is 0.002. (The increased significance of the variation in the higher-energy light-curve is in spite of the poorer statistics in this energy range, vs. 0.2-1.0 keV.) These results indicate that the differences in the light-curves of the ACIS and HRC observations may be explained by the differences in the relative sensitivity with energy of the two detectors.

Figure \ref{fig:2015enph} shows the scatter plot of photon energy vs. photon phase for the 2014--15 observations. The declining effective area of ACIS over the mission produced the reduced number of low energy photons in Fig.~\ref{fig:2015enph}, compared to Fig.~\ref{fig:2002enph}. Figs.~\ref{fig:2015_low}, \ref{fig:2015_high} show the low  and high energy light-curves for the 2014--15 observations. The 2014--15 observations show a less dramatic difference in the variability characteristics of low and high-energy photons, largely because the lowest-energy photons are simply missing in 2014--15 (compare Fig.~\ref{fig:2015enph} and Fig.~\ref{fig:2002enph}). The high energy light-curves of 2002 and 2014--15 observations are similar.

\section{X-Ray Spectrum}
\label{sec: spectra}
\begin{figure}
 \centering
 \includegraphics[width=0.45\textwidth]{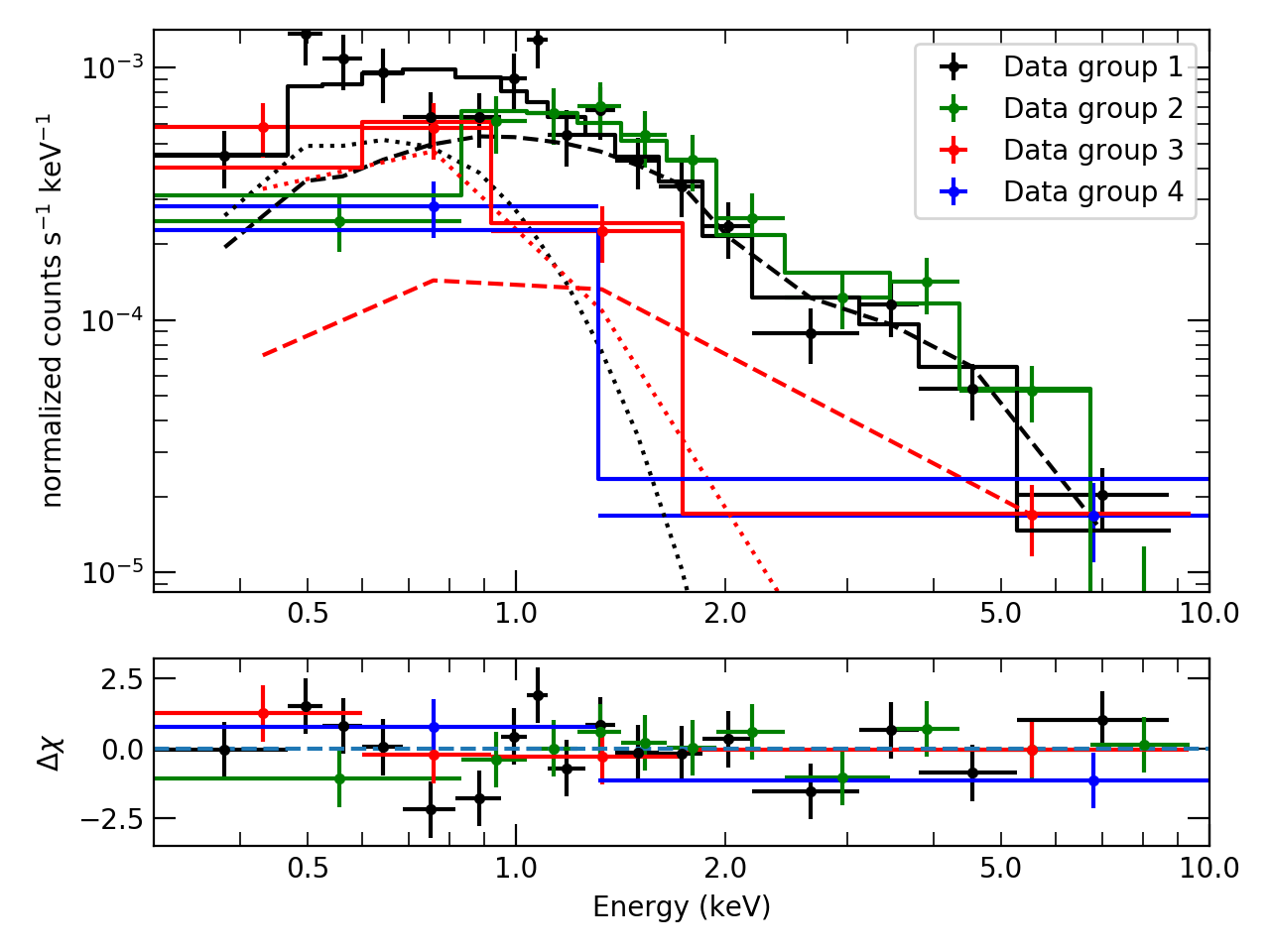}
 \includegraphics[width=0.45\textwidth]{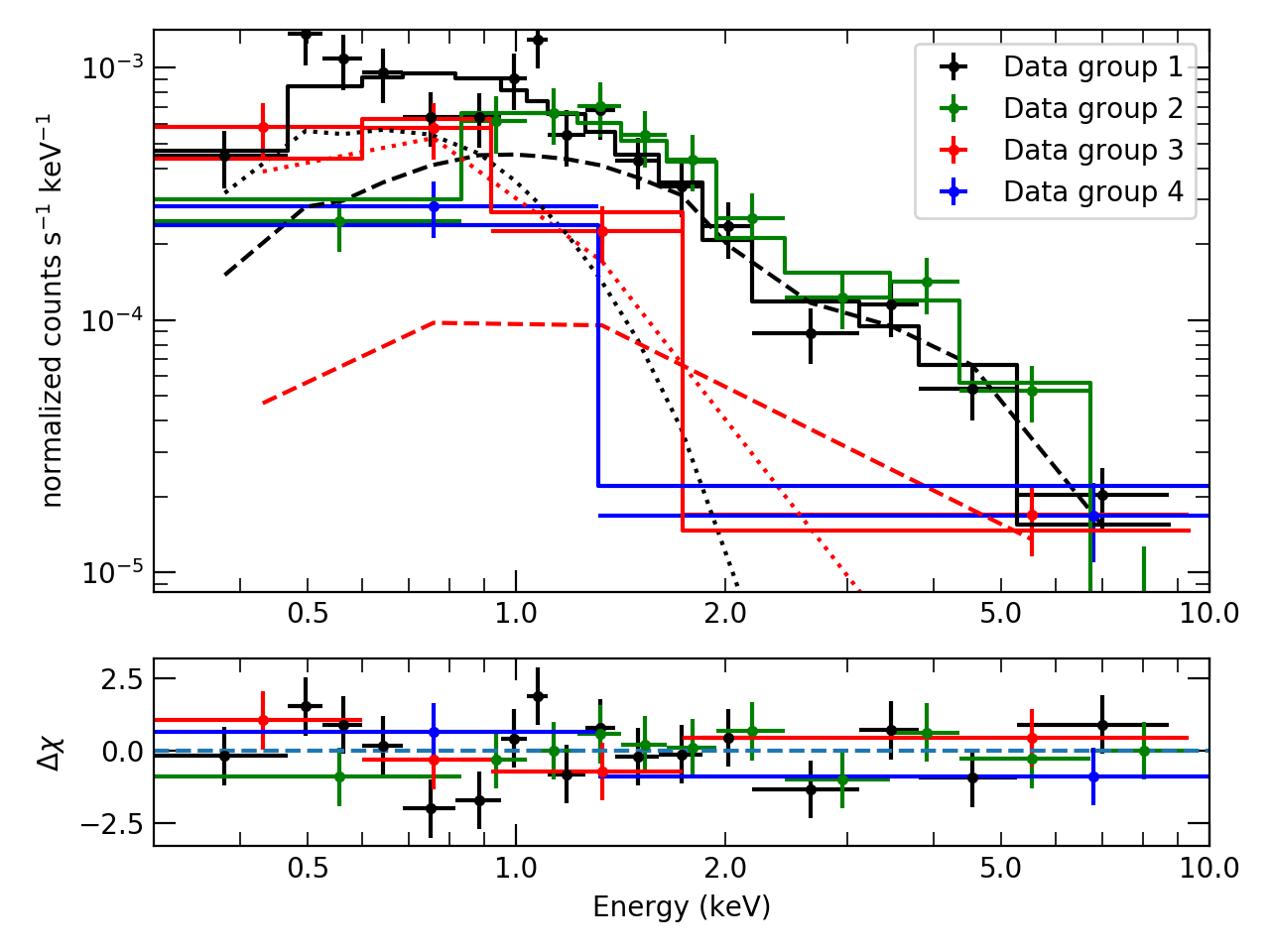}
 \caption{Spectral fits to the 2002 data (black: bright, phases -0.0-0.1, 0.4-1.0; red: faint, phases - 0.1-0.4), and 2014--15 data (green: bright, phases phases - 0.0-0.1, 0.4-1.0. blue: faint, phases  0.1-0.4). {\it Top}: Model with  power-law and blackbody components, and \texttt{tbabs} absorption. {\it  Bottom}: Model with power-law and neutron star atmosphere components, plus absorption. We also plot the thermal (dotted) and non-thermal (dashed) components of the spectrum separately for the 2002 bright (black) and faint (red) spectra.}
 \label{fig:spec_fit}
\end{figure}

We analysed the X-ray spectrum for 47 Tuc W using the 0.3 - 8.0 keV photons where the ACIS instrument has the highest sensitivity. We divided the data into 4 groups to look for changes in the spectra during eclipse, and between the 2002 and 2014--15 observations:
\begin{itemize}
 \item \textbf{D1} - Data from 2002 observations extracted from phases 0.0-0.1, and 0.4- 1.0 (higher flux).
 \item \textbf{D2} - Data from 2014--15 observations extracted from phases 0.0-0.1, and 0.4- 1.0 (higher flux).
 \item \textbf{D3} - Data from 2002 observations extracted from phases 0.1-0.4 (lower flux).
 \item \textbf{D4} - Data from 2014--15 observations extracted from phases 0.1-0.4 (lower flux).
\end{itemize}

Due to the low photon counts, we grouped the data to have a minimum of 1 photon per bin, and used C-statistics to fit the models. We used the \texttt{tbabs} model in XSPEC to model the absorption due to the interstellar medium.  We fixed the hydrogen column density towards 47 Tuc to be $3.5 \times 10^{20}$ cm$^{-2}$ \citep{bogdanov2016}, using \texttt{wilms} abundances \citep{wilms2000}. We describe the various spectral models used and the approximations made in the paragraphs below. The results of our spectral analysis are summarised in Table~\ref{table:spec_analys}. We quote 90\% confidence errorbars.

We first fit all four data groups independently with a power law (model 1). Due to the low photon counts, the parameters of the spectral fits for D3 and D4 were essentially unconstrained. We observed that the photon indices of D1 and D2 were consistent within their 90\% confidence errors, as were those of D3 and D4. Therefore, we linked the photon indices of (D1, D2) and (D3, D4) to be equal. Fitting a pegged power law, with linked photon indices (model 2), gave $\Gamma = 1.49 \pm 0.12$ for D1 and D2 (the bright phases) and $\Gamma = 2.31 \pm 0.32$ for D3 and D4 (the faint phases). 

The large change in the photon index during the flux dips suggests the possibility of two spectral components in 47 Tuc W. The hard spectrum could be emitted from the intra-binary shock (IBS). The decrease in  flux from this hard spectral component could be due to eclipsing by the companion, or due to Doppler beaming of radiation from the shocked material away from us. The softer spectrum could be thermal emission from the NS, which  would not be eclipsed by the companion except at very high inclinations (and then only for a very short phase interval).  The soft spectrum from the NS could be fit by either a blackbody (BB) (model 3) or a neutron star atmosphere (\texttt{nsatmos}, \citealt{Heinke06}) model (model 4). We assumed the emission from the neutron star to be constant with phase, as the inclination is unlikely to be large enough for a direct occultation of the NS by the small secondary (which would be very short even if it occurred, $<$5\% of the orbit). Therefore, we assume that the parameters of the softer spectrum are constant across all the data groups. Note that since there could be some X-ray emission from the IBS, even during the phases where flux decreases, we do not fix the power-law flux to be zero in data groups D3 and D4.

We modelled the BB emission using \texttt{bbodyrad} in XSPEC, limiting the BB temperature to $<$0.3 keV (to avoid the fit increasing the BB temperature to fit the high-energy component). We kept the BB component parameters identical across the 4 data groups, and allowed only the power-law component normalisation to vary between the data groups. Including the thermal component decreases the power-law index slightly  to $\Gamma = 1.16_{-0.26}^{+0.24}$. The BB component has an effective temperature $T_{eff} = 1.84_{-0.58}^{+0.48} \times 10^{6}$ K, and a radius $R_{eff} = 0.24_{-0.09}^{+0.27}$ km - consistent with emission from heated polar caps of a NS, as seen in other MSPs \citep[e.g.][]{Bogdanov06}.  From the best fit model (Fig.~\ref{fig:spec_fit} {\it (top)}, we see that the non-thermal flux is roughly twice the thermal flux even during the phase interval 0.1-0.4. 

We also fit the soft component with a neutron star atmosphere, using the \texttt{nsatmos} model of XSPEC. We fix the neutron star mass to 1.4 M$_{\odot}$, the neutron star radius to 10 km, and the distance to 4.53 kpc \citep{bogdanov2016}. The \texttt{norm} parameter in this model allows a rough estimate of the fractional part of the neutron star emitting, which we left free, but tied to the same value between the data groups (model 4). The radius of the region emitting thermal radiation is much smaller than the NS radius, indicating the presence of hot spots on the NS. Fixing the \texttt{norm} to 1 and varying the radius of the NS gives an extremely small radius ($\sim$ 5 km), which is not plausible.

The best fitting model shown in Fig.~\ref{fig:spec_fit} { (bottom)}, has a photon index $\Gamma = 1.04_{-0.27}^{+0.26}$, similar to that in the \texttt{tbabs*(pegpwrlw+bbodyrad)} model. We find that the temperature and radius of the hotspot on the NS are $1.05_{-0.37}^{+0.36} \times 10^6$ K and $1.29_{-0.49}^{+1.61}$ km, respectively.  These are again consistent with the values for NS hotspots in other MSPs \citep[e.g.][]{Bogdanov06}. Since \texttt{nsatmos} is a more physically motivated model for emission from a neutron star surface, we use this model for further analysis. 

While fits to both BB and NS atmosphere models indicate the presence of small hotspots, we do not directly detect pulsations to prove the expected rotational variability. This is not surprising, as gravitational bending of light is theoretically predicted to suppress pulsed hotspot fractions to $\lesssim$50\% \citep{Pechenick83,Beloborodov02,Bogdanov08}. The pulsation fraction of thermal hotspot emission will vary due to the relative angles between the observer's line of sight, the spin axis, and the position of hotspots \citep[see e.g.][]{Bogdanov08}. Empirical observations of pulsations indeed find broad sinusoidal pulses and modest pulsed fractions  \citep[e.g.][]{Becker93,Pavlov97,cameron2007,Guillot19}. In particular, the 47 Tuc HRC-S  observations did not detect pulsations from the majority (11 of 19, including 47 Tuc W) of MSPs in 47 Tuc \citep{cameron2007}. Although 47 Tuc W is the brightest of the 47 Tuc MSPs, we attribute only $\sim$45\% of its HRC-S photons  to thermal emission, with the rest to (presumably unpulsed) shock emission, which would therefore dilute hotspot pulsations.

\begin{table*}
\centering
\caption{Summary of spectral analysis of 47 Tuc W}
\begin{tabular}{cccccc}
\hline
Data Groups         & Model parameters & \texttt{tbabs*pegpw}                                          & \texttt{tbabs*pegpw}(linked indices)                                           & \texttt{tbabs*(pegpw+bbodyad)}                                           & \texttt{tbabs*(pegpw+nsatmos)}                                           \\ \hline
D1 & Power law index  & $1.60 \pm 0.15$                                    & $1.50 \pm 0.12$                                     & $1.16_{-0.26}^{+0.24}$                                    & $1.04_{-0.27}^{+0.26} $              \vspace{0.5em}                      \\ 
                    & Power law flux   & $1.07_{-0.13}^{+0.14} \times 10^{-14}$  & $1.13_{-0.13}^{+0.14} \times 10^{-14}$  & $1.04_{-0.15}^{+0.16} \times 10^{-14}$  & $1.01_{-0.16}^{0.17} \times 10^{-14}$ \vspace{0.5em}   \\
                    & T$_{eff}$             & -                                                 & -                                                 & $1.84_{-0.58}^{+0.48} \times 10^{6}$        & $1.05_{-0.37}^{+0.36} \times 10^{6}$    \vspace{0.5em}   \\
                    & Radius          & -                                                 & -                                                 & $2.38_{-0.87}^{+2.73} \times 10^{-1}$    & $1.29_{-0.49}^{+1.61}$                                 \vspace{0.5em}        \\
                    & Thermal Flux & - & - & $1.58_{-0.61}^{+0.60} \times 10^{-15}$ & $2.02_{-0.66}^{+0.62} \times 10^{-15}$  \vspace{0.5em} \\ \hline
D2 & Power law index  & $1.29 \pm 0.21$                                     & $1.50 \pm 0.12$                                     & $1.16_{-0.26}^{+0.24} $                                    & $1.04_{-0.27}^{+0.26} $                     \vspace{0.5em}                 \\
                    & Power law flux   & $1.54_{-0.22}^{+0.26} \times 10^{-14}$  & $1.42_{-0.18}^{+0.20} \times 10^{-14}$ & $1.45_{-0.21}^{+0.24} \times 10^{-14}$  & $1.45_{-0.22}^{+0.26} \times 10^{-14}$  \vspace{0.5em}  \\
                    & T$_{eff}$             & -                                                 & -                                                 & $1.84_{-0.58}^{+0.48} \times 10^{6}$        & $1.05_{-0.37}^{+0.36} \times 10^{6}$   \vspace{0.5em}   \\
                    & Radius          & -                                                 & -                                                 & $2.38_{-0.87}^{+2.73} \times 10^{-1}$    & $1.29_{-0.49}^{+1.61}$                          \vspace{0.5em}           \\
                     & Thermal Flux & - & - & $1.58_{-0.61}^{+0.60} \times 10^{-15}$ & $2.02_{-0.66}^{+0.62} \times 10^{-15}$  \vspace{0.5em} \\ \hline
D3 & Power law index  & $2.43_{-0.38}^{+0.40}$                                     & $2.30_{-0.31}^{+0.33}$                                     & $1.16_{-0.26}^{+0.24}$                                    & $1.04_{-0.27}^{+0.26} $        \vspace{0.5em}                              \\
                    & Power law flux   & $3.98_{-0.82}^{+0.98} \times 10^{-15}$ & $4.07_{-0.85}^{1.01} \times 10^{-15}$ & $3.04_{-1.38}^{+1.64} \times 10^{-15}$  & $2.41_{-1.29}^{+1.65} \times 10^{-15}$ \vspace{0.5em}   \\
                    & T$_{eff}$             & -                                                 & -                                                 & $1.84_{-0.58}^{+0.48} \times 10^{6}$        & $1.05_{-0.37}^{+0.36} \times 10^{6}$    \vspace{0.5em}   \\
                    & Radius          & -                                                 & -                                                 & $2.38_{-0.87}^{+2.73} \times 10^{-1}$    & $1.29_{-0.49}^{+1.61}$                 \vspace{0.5em}                  \\
                     & Thermal Flux & - & - & $1.58_{-0.61}^{+0.60} \times 10^{-15}$ & $2.02_{-0.66}^{+0.62} \times 10^{-15}$  \vspace{0.5em} \\ \hline
D4 & Power law index  & $2.01_{-0.56}^{+0.58}$                                       & $2.30_{-0.31}^{+0.33}$                                     & $1.16_{-0.26}^{+0.24}$                                    & $1.04_{-0.27}^{+0.26} $                \vspace{0.5em}                     \\
                    & Power law flux  & $5.03_{-1.47}^{+1.88} \times 10^{-15}$                  & $5.00_{-1.47}^{1.81} \times 10^{-15}$   & $4.38_{-2.05}^{+2.45} \times 10^{-15}$ & $3.73_{-1.92}^{2.48} \times 10^{-15}$ \vspace{0.5em}  \\
                    & T$_{eff}$             & -                                                 & -                                                 & $1.84_{-0.58}^{+0.48} \times 10^{6}$        & $1.05_{-0.37}^{+0.36} \times 10^{6}$    \vspace{0.5em}   \\
                    & Radius          & -                                                 & -                                                 & $2.38_{-0.87}^{+2.73} \times 10^{-1}$    & $1.29_{-0.49}^{+1.61}$         \vspace{0.5em}       \\   
                     & Thermal Flux & - & - & $1.58_{-0.61}^{+0.60} \times 10^{-15}$ & $2.02_{-0.66}^{+0.62} \times 10^{-15}$  \vspace{0.5em} \\ \hline   
\end{tabular}
\justify
Note: Flux is expressed in  units of ergs/cm$^2$/s (between 0.3-8.0 keV), $T_{eff}$ in K and radius in km. D1 - Photons from 2002 observations outside of dips. D2 - Photons from 2014--15 observations outside of dips. D3 - Photons from 2002 observations during dips. D4 - Photons from 2014--15 observations during dips. "Radius" refers to an effective radius of the region emitting thermal radiation.  Error-bars represent 90\% confidence intervals, i.e. $1.65\sigma$.
\label{table:spec_analys}
\end{table*}

From Table \ref{table:spec_analys}, we see that the thermal flux is much smaller ($\approx 1/5$) than the non-thermal one outside dips. The non-thermal and thermal luminosities are estimated as $L_{X, NT} = (2.5 \pm 0.4) \times 10^{31}$ ergs/s, and $L_{X, NS} = 5_{-2}^{+1} \times 10^{30}$ ergs/s between 0.3 - 8.0 keV for the 2002 data. The non-thermal flux seems to have increased during the 2014--15 observations, compared to the 2002 observations. We verify this change in the power-law flux using two methods --- First, we tie the power-law flux of (D1, D2) and (D3, D4), and check the change in C-statistic value. This increases the C-statistic by 9.07 while decreasing the d.o.f by 2, i.e. the p-value $= 0.01$ for constant flux. Next, we use the model \texttt{const*pegpwrlw} and fix the constant to 1.0 for D1. We also tie the power-law flux of all data groups. Fitting this model results in a constant value of $1.4_{-0.2}^{+0.4}$ for data group D2, i.e. the ratio of fluxes $> 1.0$ with $\approx 3.3\sigma$ confidence for data group D2 compared to D1.

\subsection{Verifying the HRC light-curve}
We use our best fitting spectral model for the ACIS-S data to check if {\it Chandra} HRC-S should have detected variability, if 47 Tuc W's spectrum and variation had remained the same. We use the {\it Chandra} PIMMS tool\footnote{http://cxc.harvard.edu/toolkit/pimms.jsp} to estimate the background and the source count rates for the thermal and non-thermal X-ray emission separately. We use the \texttt{tbabs*(bbodyrad+pegpw)} model due to the limited capability of the PIMMS tool. We further use the Chandra Ray Tracer \citep[ChaRT;][]{Carter2003} tool and \texttt{simulate\_psf} function in CIAO to calculate the fraction of the PSF within a $1\arcsec$ region. 

We find the estimated count rates for the power-law and BB components to be $3.1 \times 10^{-4}$ cts/s and $1.8 \times 10^{-4}$ cts/s, respectively. The count rate from the power-law flux should decrease to $9.0 \times 10^{-5}$ cts/s during the dip. The background count rate of HRC is $2.4 \times 10^{-4}$ cts/s for an extraction region of $1 \arcsec$. The total exposure of all 2014--15 observations is 786.22 ks. Thus each bin in our light-curve (of bin size 0.1) corresponds to $\sim 79$ ks. The estimated number of counts from the non-thermal power-law in each HRC bin is $24.2$ during phases 0.0--0.1, 0.4--1.0, and $7.0$ during phases 0.1--0.4. The estimated count-rate from thermal black-body flux is $13.9$ per bin. Each bin also should have $18.9$ background counts (subtracted off). Assuming Gaussian errors, the total net source counts per HRC bin during phases 0.0-0.1 \& 0.4-10.0 would be $38.1 \pm 7.5$, and in phases 0.1 - 0.4 would be $20.9 \pm 6.3$. Comparing our results to Fig.~\ref{fig:lc_2005}, we observe that while the maximum count-rate during phases 0.0--0.1 \& 0.7--1.0 ($\sim 40$ counts per bin) is approximately equal to that expected, the minimum count-rate during phases 0.1-0.4 (corresponding to the dip in ACIS light-curves; the count-rate is consistently $\geq 25$ counts per bin) is higher than the expected $20.9 \pm 6.3$, i.e., while we expect a total of $62.7 \pm 10.9$ between phases 0.1 \& 0.4, we observe $\sim 82.5$ net counts. Additionally, phases 0.4--0.7 also show a similar count-rate to 0.1--0.4. Thus, the intra-binary shock in the 47 Tuc W system could have intrinsically changed between the 2002 ACIS and the 2005 HRC observations.

\section{Comparison between intra-binary shock models and X-ray and optical data}
\label{sec:optical}
In this section, we compile optical and X-ray light-curves, describe our intra-binary shock model, and explain results from our fitting.

\subsection{X-ray and optical data}

To model the non-thermal peaks of IBS emission we created a  1.5-10\,keV ACIS light-curve. Given that a non-binned KS test shows no significant difference in the phase structure of the 1.5-10\,keV counts for the two epochs ((KS test value 0.15, $p$-value = 0.14), we combined the 2002 and 2015 data sets, folding on the orbital period of \citet{ridolfi2016}.  The X-ray maximum is broad, covering over half of the orbit, $\phi_B \approx 0.45 - 0.10$, with a hint of a double peak structure ($\phi_{p,1} \approx 0.58$, $\phi_{p,2} \approx 0.93$) bracketing the optical maximum (Fig.~\ref{fig:fitting}).  The radio eclipse covers $\phi_{\rm{B}} = 0.09 -- 0.43$ \citep{ridolfi2016}. 

We converted the normalised  optical fluxes and errors in \citet{bogdanov2005} and the ultraviolet magnitudes and errors in \citet{cadelano2015} to units of erg/cm$^2$/s/Hz ($f_{\nu}$) and plot them in Fig.~\ref{fig:fitting} (two periods are shown for clarity). As expected for a cool, heated companion, the optical light-curves show relatively shallow dips, while the UV light-curves show a deeper modulation from the higher temperature of the heated part of the companion near the $L_1$ point. 

All optical/UV light-curves show a maximum near $\phi_{\rm{B}}\sim 0.75$ ($\phi_B=0$ denotes the phase of the pulsar ascending node), indicating strong heating of the pulsar companion. Several light-curves appear slightly asymmetric, with the UV curves delayed; F300X appears to show a maximum at $\phi_{\rm{B}}\approx 0.8$. It is unclear if the offset is significant, but if real it may indicate asymmetric local heating as inferred for other redback-type binaries. For example, IBS-accelerated particles may contribute to the heating, being ``ducted'' by companion fields to the surface and producing heated magnetic poles \citep{sanchez2017}; this would most strongly affect the UV light-curves. Similarly, asymmetries in the light-curves could also arise from atmospheric circulations on the surface of the companion \citep{Kandel_2020}.

\subsection{Intra-binary shock modelling}
\label{sec: IBS}

To model the companion heating, we use the ICARUS code of \citet{breton2012}. This requires tables of  stellar atmosphere colours as a function of local surface temperature and gravity. We extract these from the Spanish Virtual Observatory fold of the BT-Settl atmosphere models \citep{allard2012} through the responses of the individual {\it HST} filters, convert the normalised surface fluxes to $f_\nu$, and supply them to the ICARUS code. The four optical bands show shallow modulation. The five F390W points are puzzling, with our models over-predicting this flux by a factor of $2\times$, although the shape is as expected. We suspect that there is a difference between the calibration of \citet{cadelano2015}, and our filter assumptions, although we have not been able to identify a specific error. 
The F300X fluxes follow the expected light-curve quite closely. To deal with this we adopt a simple offset in the F390W magnitudes.

The non-thermal X-ray emission is attributed to synchrotron emission \citep{bogdanov2005,wadiasingh2018} from particles accelerated by the intra-binary shock formed by the interaction of the pulsar wind and the stellar wind from the main-sequence companion. To study the direct X-ray emission from the IBS, we have further extended the ICARUS IBS code \citep{romani2016} with a module which follows the variation of the synchrotron emission across the shock \citep{Kandel19}.  In this model, the shape of the IBS is controlled by the wind momentum ratio $\beta$ and a wind velocity asymmetry factor $f_v$.  The IBS spectral behavior is controlled by the evolution of the bulk flow speed $\Gamma$, the spectral index of the accelerated electron power-law, and the characteristic magnetic field strength at the IBS nose  (the point of the shock on the line connecting the two stars). 

The X-ray light-curves of many redback MSPs (such as 47 Tuc W) show peaks around pulsar superior conjunction. In the standard modelling scenario, this indicates that the momentum of the massive companion wind dominates over the momentum of the pulsar wind, $\beta>1$, sweeping the shock back to wrap around the pulsar \citep{romani2016,wadiasingh2018,alnoori2018}. For our modelling we assume, following \citet{Kandel19}, that the shocked pulsar wind accelerates adiabatically along the IBS contact discontinuity
\begin{equation}
\Gamma \approx 1.1 \left(1+0.2 \frac{s}{r_0}\right)~,
\end{equation}
where $r_0$ is the stand-off distance of the shock from the nose, and $s$ is the arc length from the nose of the IBS to the position of interest along the IBS. We also assume that the magnetic field at the nose of the IBS is 10 G.  

For many black widows and redbacks, the IBS synchrotron emission dominates over non-thermal X-rays directly from the pulsar. Above 1.5 keV, we expect little to no emission from the NS surface, so we treat the 1.5-10\, keV X-rays as produced entirely by the shock. For 47 Tuc W, the relatively hard observed X-ray spectrum (photon index  $\sim 1.04$) implies negligible contribution by the IBS emission to the flux in the optical/UV bands, which are dominated by the pulsar-heated face of the companion. 

\begin{figure}
\centering
\includegraphics[width=0.48\textwidth]{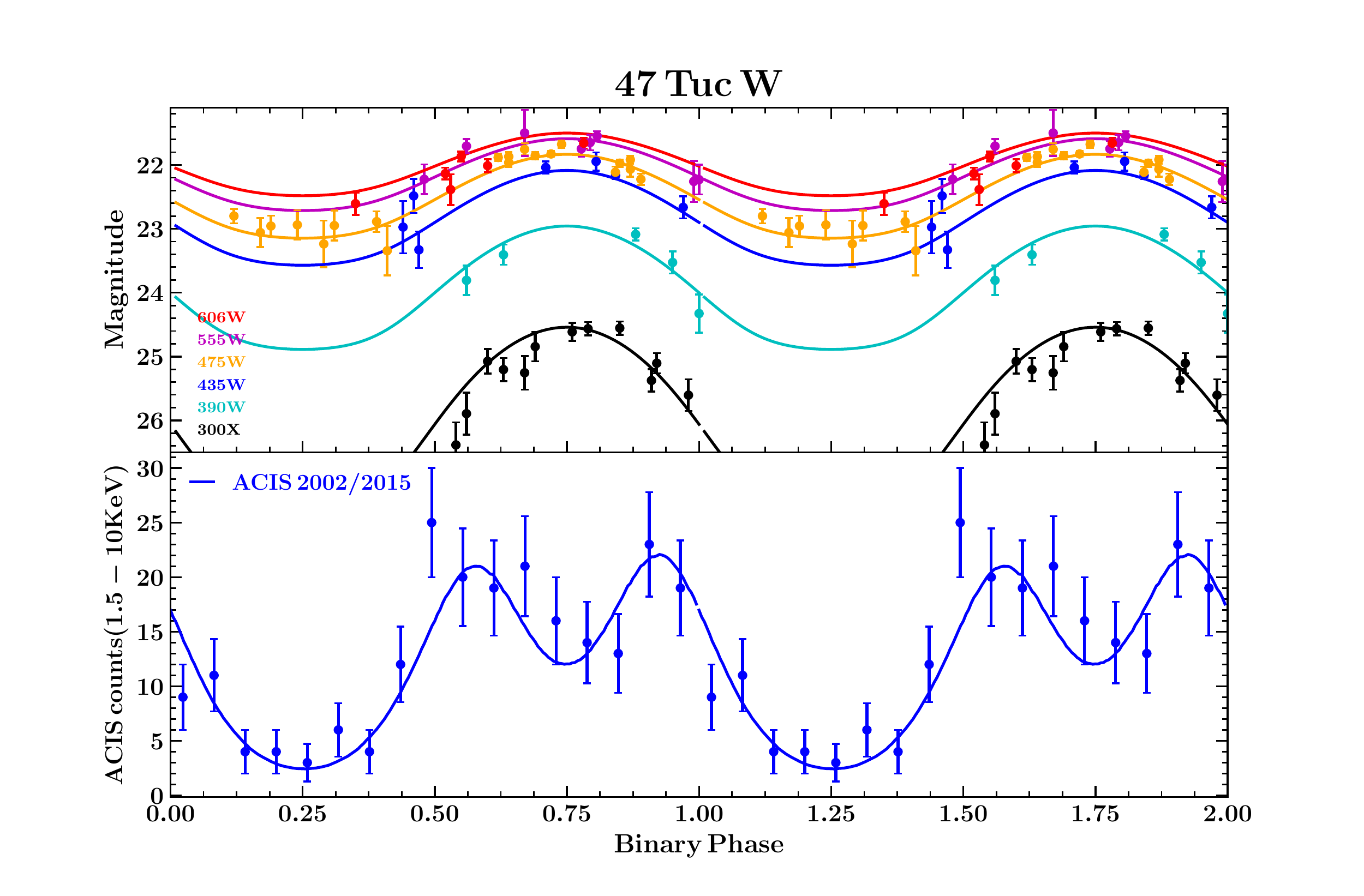}
\caption{Above: {HST} ACS and WFC3 measurements of 47 Tuc W in four optical and two UV-dominated bands (after \citealt{bogdanov2005} and \citealt{cadelano2015}). Our best-fitting model light-curves are superimposed. Note that the F390W points, shown in cyan, are shifted by a phase-independent offset fit as $\Delta m =0.4\pm0.05$ mag to deal with an apparent calibration error.
Below: 1.5-10\,keV X-ray light-curve from the combined 2002 and 2014--15 data sets, with the model X-ray light-curve for the geometry of the combined fit.}
\label{fig:fitting}
\end{figure}

To constrain the system parameters, we have fitted the optical and X-ray data independently. The optical model is sensitive to the binary inclination $i$, while the X-ray light-curve is sensitive to $i$ and the shock geometry. Recall that the F390W points are shifted by a phase-independent offset fit as $\Delta m =0.4\pm0.05$ mag to deal with an apparent calibration error. Fig~\ref{fig:fitting} shows the model-fitted optical and X-ray light-curves, and the fitted parameters are listed in Table \ref{tab:fit_param}. Note that the UV phase shift, if real, might be due to particle precipitation to magnetic poles \citep{sanchez2017}, or  atmospheric circulation on the companion \citep{Kandel_2020}. However, higher S/N light-curves are needed to motivate such detailed modelling.

The fitted properties in Table \ref{tab:fit_param} seem generally reasonable for a redback MSP. The inclinations inferred from both X-ray and optical modelling are consistent with each other, as are the heating luminosities. However, given the poor data quality, the best fit values have large uncertainties. The inferred direct heating luminosity (as in \citealt{romani2016}), $L_H \sim 6\times10^{33}$ erg/s, is of order $10\%$ of the  spindown luminosity of similar redback MSPs of $\sim 6\times10^{34}$ erg/s (e.g. \citealt{archibald2009}). (The intrinsic $\dot{P}$ of 47 Tuc W is not known, since it is accelerated in the gravitational potential of its globular cluster; using the absolute value of the measured $\dot{P}$ of \citealt{ridolfi2016} gives a maximum spindown luminosity of $2\times10^{35}$ erg/s.) For an assumed pulsar mass of 1.4 M$_{\odot}$, the inferred companion mass is 0.16 M$_{\odot}$.  Constraining the mass-loss rate of the companion would require knowledge of the intrinsic spin-down luminosity of the pulsar ($\dot{E}$), the orbital velocity of the companion ($v_{orb}$), and the ratio of wind speed to orbital speed of the companion ($f_v = v_w/v_{orb}$) in addition to the wind momentum flux ratio ($\beta = \dot{M}v_w c/\dot{E}$) \citep{romani2016}. Assuming a pulsar mass of 1.4 M$_{\odot}$, $f_v > 1$, and $\dot{E} \approx 10^{34}$ erg/s, we find $\dot{M} < 6 \times 10^{-10} M_{\odot}$/yr, which is not highly constraining.

Looking at the X-ray light-curve in Fig.~\ref{fig:fitting}, there seems to be marginal evidence that the two peaks are asymmetric. Such asymmetry could be due to sweepback caused by the finite speed of the companion wind, characterized by $f_v \equiv v_{\rm{w}}/v_{\rm{orb}}$, where $v_{\rm{w}}$ is the speed of companion wind and $v_{\rm{orb}}$ is the orbital speed. In general, the smaller $f_v$, the larger the X-ray peak asymmetry. Because of the poor data quality, we only quote a lower-bound, estimated by finding the  $f_v$ for which $\chi^2$ increases to the value found at the 1$\sigma$ $i$ bounds. With this definition, we infer that $f_v\gtrsim 1$. Higher S/N data would be necessary for a more precise constraint. 

\begin{table*}
\centering
\begin{tabular}{l c c c}  
\hline  
Parameter & Symbol & Optical Value & X-ray value\\
\hline
Inclination (degrees) & $i$ & $67.6 \pm 12.0$ & $51.64 \pm  9.99$\\
Roche lobe filling factor & $f_1$ & $0.65\pm 0.03$ & - \\
Star temperature (night, K) & $T_N$ & $4973 \pm 73$ & - \\
Heating luminosity (erg s$^{-1}$)& $L_H$ & $(5.74\pm 0.44)\times 10^{33}$ & $(3.45\pm 2.05)\times 10^{33}$\\
Wind momentum ratio & $\beta$ & -&$5.1\pm 0.8$\\
Chi-square & $\chi^2/\nu$ & $104/53$ & $14/12$\\  \hline
\end{tabular}
\caption{Parameters for fit of IBS to optical \& X-ray light-curves.}
\label{tab:fit_param}
\end{table*}

\section{Conclusions}
\label{sec:conclusions}

In this work, we analyse new and archival {\it Chandra} observations of 47 Tuc between  2000 and 2015 to test whether the redback pulsar 47 Tuc W shows signs of changes in its X-ray properties, which might signal a transitional MSP. The ACIS-S X-ray light-curves of 47 Tuc W in the years 2002 and 2014--15 show a drop in the count rate during the phases $\sim 0.1 - 0.4$ (phase 0 corresponds to time of periastron passage). However, the X-ray light-curve of 2005--06 doesn't show a statistically significant modulation in counts per bin during this phase range. We see that the orbital modulation in the count rate is more pronounced at higher energies. Thus instruments like {\it Chandra} HRC-S, which are primarily sensitive to soft X-rays, cannot clearly detect such flux modulations. We do see some evidence for changes in the X-ray light-curve from 2002 to 2005 (where the predicted count rates in HRC-S data are marginally different from expectations based on the 2002 data) and 2014--15 (where the light-curve shape is subtly different). However, the significant differences in the detector, or detector contamination, between epochs make this difficult to verify. Should such changes be present, they would likely signify changes in the wind from the companion. 

Our analysis of the X-ray spectrum shows that the X-ray emission consists of two components - a dominant hard non-thermal power-law ($\Gamma \sim 1.1$) which shows orbital modulation (drops by a factor of 3 during the phases 0.1-0.4) and a soft thermal component, which is consistent with being constant at all phases. The soft emission can be fit using a thermal BB with $T_{BB} = 1.8_{-0.6}^{+0.5} \times 10^{6}$ K, or a NS H atmosphere with $T_{eff} = (1.0 \pm 0.4) \times 10^{6}$K. The unabsorbed non-thermal luminosity, $L_{X, NT} = (2.5 \pm 0.4) \times 10^{31}$ ergs s$^{-1}$ is about 5 times the thermal luminosity, $L_{X, NS} = 5_{-2}^{+1} \times10^{30}$ ergs s$^{-1}$. The thermal X-ray emission is consistent with arising from a hydrogen atmosphere hot spot, of radius 0.5-3 km, in agreement with the analysis of \citet{Bogdanov06}. We use an updated pulsar ephemeris \citep{ridolfi2016} to search for spin modulation of the X-rays in the 2005-06 HRC data, but do not detect X-ray pulsations.

The X-ray orbital modulation, and the companion star-dominated orbital light-curve in the optical and UV, allow us to study the properties of the IBS and the geometry of the binary system. Modelling the double peaked X-ray light-curve, and the optical light-curve, using the ICARUS IBS code indicates that the companion is heated by the pulsar spindown power.
Though the companion doesn't fill its Roche lobe, the heating of the companion leads to a strong stellar wind which has $\sim 6\times$ the momentum of the pulsar wind; i.e., the IBS is wrapped around the pulsar.
The apparent offset of the peak of the F300X light-curve might be attributed to magnetic pole hot spots caused by channelling of pulsar wind particles by the companion field. Higher signal-to-noise ultraviolet light-curves (such as those of XSS J12270-4859; \citealt{rivera2018mid}) are needed to model such effects.

Our temporal and spectral analyses are limited by the low photon counts from the source. Longer exposure or use of instruments like Lynx and/or LUVOIR, with larger effective area but similar angular resolution, would allow us to study this binary system in more detail. Higher X-ray photon count rates would allow us to study the double peaked nature of the light-curve in more detail and constrain the properties of the IBS. More photons in the region corresponding to radio eclipse would also allow better modelling of the NS emission. Given the cluster crowding in the optical/UV, space observations with e.g. the planned LUVOIR mission would be needed. Happily, because of the many interesting targets in 47 Tuc, there is hope for serendipitous 47 Tuc W studies from future observations of this important cluster.

\section*{Data Availability Statement}
The X-ray data underlying this article are available in the Chandra Data Archive and can be accessed using the Chandra Search and Retrieval (ChaSeR; \href{https://cda.harvard.edu/chaser/}{https://cda.harvard.edu/chaser/}) tool. The {\it Hubble Space Telescope} optical observations corresponding to the optical data used in this article are available at the Mikulski Archive for Space Telescopes (MAST; \href{http://archive.stsci.edu/hst/}{http://archive.stsci.edu/hst/}). The light-curves used in this article are presented in \citet{bogdanov2005} and \citet{cadelano2015}.

\section*{Acknowledgements}

We thank A. Ridolfi for giving insights into the evolution of the binary orbit of 47 Tuc W. 
COH acknowledges support from NSERC Discovery Grant RGPIN-2016-04602, a Discovery Accelerator Supplement, and the University of Alberta Research Experience program. DK and RWR were supported in part by NASA grants 80NSSC17K0024 and 80NSSC18K1712.




\bibliographystyle{mnras}
\bibliography{47TucW_paper} 





\bsp	
\label{lastpage}
\end{document}